# Universal programmable on-chip metasurface building blocks for arbitrary high-order mode manipulation

Jinlong Xiang[†], Zhiyuan Tao[†], Xuhan Guo[*], Yong Zhang, Yaotian Zhao, and Yikai Su

*State Key Laboratory of Advanced Optical Communication Systems and Networks, Department of Electronic Engineering, Shanghai Jiao Tong University, Shanghai 200240, China*
*† These authors contributed equally to this work.*
*\*Corresponding author: guoxuhan@sjtu.edu.cn*

**Abstract:** On-chip mode-division multiplexing (MDM) has been emerging as a promising technology to further enhance the link capacity and bandwidth of data communications with multiple mode channels. Both mode converters and mode exchangers are indispensable fundamental components for flexible mode operations. While several configurations have been developed previously, it is still very challenging to efficiently manipulate arbitrary high-order modes in a versatile way to reduce the R&D and prototyping costs. Here we initiate a breakthrough with a simple yet universal generic mode operator building block concept utilizing metasurface structures. The programmable arbitrary high-order mode operators can realize mode conversion and exchange simultaneously with competitive and uniform performance, high stability and compact footprints, offering a quintessential step change for on-chip multimode optical interconnections.

Advanced multiplexing technologies have been playing extremely important roles to satisfy the continuously increasing demands for high capacity of data communication [1-4]. MDM technology, which leverages the extra degree of freedom of orthogonal spatial modes in multimode waveguides, is emerging to boost significantly the scalability and flexibility of both the transmission capacity and bandwidth [5, 6]. Benefiting from its footprint compactness and compatibility with the complementary metal-oxide-semiconductor (CMOS) fabrication process, silicon on-chip MDM system has attracted extensive attention. Various MDM devices have been demonstrated on a silicon-on-insulator (SOI) platform, such as mode (de)multiplexers [7-11], mode switches [12-14], multimode crossings [15, 16], multimode waveguide bends [17, 18] and reconfigurable multimode silicon photonic integrated circuits (PICs) [19, 20]. Among them, a mode converter and a mode exchanger are two of the most important and basic functional elements for flexible and efficient on-chip mode manipulation. Mode converters can transform given modes into desired modes, thus laying the foundation of parallel information communication. While mode exchangers can enable the data exchange between different mode channels, thus utilizing network resources more efficiently and facilitating flexible network performance.

Previously, significant efforts have been devoted to realize efficient on-chip mode conversions. The working principles of these devices can be generally categorized into the following four types [21]. (i) Phase matching method: The basic idea is to choose the appropriate waveguide width to guarantee that the mode in the access waveguide has the same effective refractive index with the desired mode in the bus waveguide. Considerable mode converters and mode (de)multiplexers implemented with asymmetry directional couplers (ADCs) have been reported [7-10] based on this principle. To further improve the fabrication tolerance thus scaling up the mode channel numbers, subwavelength grating structures (SWG) have been proposed [22-24]. Besides, there are other resonant coupling devices including both directional grating couplers and contra-directional grating assisted couplers [25-27]. (ii) Constructive interference of coherent scattering: The input mode will evolve into various high-order modes which interfere with each other in the conversion region. Consequently, the desired mode profile can be formed at the output port. Traditionally, multimode interference (MMI) couplers [28-30], photonic crystal waveguides [31, 32] or cascaded tapers [33] are employed. Moreover, several computer-generated nanostructures based on inverse-design algorithms have also been demonstrated [34-37]. (iii) Beam forming technique: The concept follows that the $n_{th}$ ($n \geq 1$) order *TE* mode can be treated as a combination of $n+1$ antiphase adjacent $TE_0$-like modes. To obtain the $TE_n$-$TE_0$ mode converter, the $n+1$ antiphase components of the $TE_n$ mode should travel through different effective lengths in the conversion field, thus achieving the same phase and eventually to combine into the $TE_0$ mode. Typically, Mach-Zehnder interferometer (MZI) structures are utilized, given that the phase differences between their arms can be flexibly tuned [38, 39]. (iv) Metasurface: Recently, quite a few mode converters based on metasurface structures have been proposed for both the TE polarization [40-45] and the TM polarization [46, 47]. The mode conversion between two specific modes is achieved by imposing specific refractive-index perturbations on a silicon waveguide.

Compared with the abundant demonstrations of on-chip mode conversions, only a few mode exchangers have been proposed [48-55]. Since the mode exchangers are built on the basis of mode converters, their working principles can be classified into exactly the same four techniques discussed above. More specifically, ADCs [48-50] and microring resonators (MRRs) [52, 53] are employed based on the phase matching method; Utilizing the inverse-design approach, the mode exchange between $TE_0$ and $TE_1$ modes is realized [54]; Besides, the straightforward beam shaping method is used to implement the $TE_0$-$TE_1$ mode exchanger with MZI [51]; Recently, metasurface is also proposed for on-chip mode exchangers [55].

An ideal mode converter or mode exchanger should have the merits of mode manipulation with low insertion loss (IL), low crosstalk, broad bandwidth, compact footprints, and large fabrication tolerance. On-chip MDM systems, where particular modes are allocated to optimally perform desired functionalities, however, currently require considerable designs and optimization iterations for different mode operators, leading to long development times [21]. It should be noticed that for method (i), the different high-order modes inevitably incur different designs to satisfy the phase matching conditions, and it will become almost impossible for both ADC and SWG structures to be extended for mode order higher than 15 ($n>15$) mode operators, owing to the large effective refractive index contrast between two concerned modes. Arbitrary high-order mode operators, in principle, can be obtained with the last three strategies, but all with compromised performance. (ii-iii) usually come with large footprints. Although ultra-compact mode operators have been reported with decent performance utilizing the inverse design methodology, its computation time will drastically increase when scaled up to higher-order mode ($n>3$). Meanwhile, the irregular nanostructures normally demand fabrication techniques with high accuracy, making it less feasible in real applications. (iv) represents the most compact devices reported by far, but higher-order mode operators ($n>5$) haven't been experimentally demonstrated. Good simulation results have been reported to realize arbitrary mode-order conversion [45]. However, the conversion efficiency will decrease dramatically for high-order mode (e.g., 94.5% for the $TE_1$ mode while 82.5% for the $TE_2$ mode even in theory at the wavelength of 1550 nm). Besides, introducing the half-etched $Si_3N_4$ slots on silicon waveguides will definitely increase the fabrication complexity. Although the novel approach is claimed to be universal for on-chip mode operators [55], the design actually is mode-specific and the involved parameter space in optimization process will grow rapidly as the increase of concerned mode order. Moreover, the geometry of metamaterial structures will get much smaller for higher-order modes, thus further posing challenges on fabrication techniques. Therefore, a simple solution to unlimited high-order mode operators with constant good performance, compact footprints, high scalability and feasibility in practice for efficient mode manipulation is extremely desirable.

In this paper, we propose for the first time, to the best of our knowledge, a universal scheme to program arbitrary high-order mode operators in an easy yet generic building block approach. This method is inspired by electronic Field Programmable Gate Arrays (FPGA), where any arbitrary high-order mode operators can be implemented by a simple topology consisting of basic low-order mode operators to realize different multimode functionalities through programming. We first introduce our design philosophy with the primitive $TE_0$-$TE_2$ mode operator exploiting fully etched dielectric metasurface slots in the SOI platform, which will function as the basic building block for high-order mode operators. The simulated insertion loss is less than 1 dB and the crosstalk are lower than -10 dB across the wavelength band from 1500 nm to 1600 nm. The operating principles are theoretically analyzed using the beam shaping technique together with the coupled mode theory (CMT). After that, any even-order mode operators can be acquired by simply programming parallel arrangement of a set of building blocks with an appropriate waveguide width. Employing the same approach, any odd-order mode operators can be realized by directly shifting and chopping part of the even-order mode operators with a fixed width. We emphasize that the performance of all the programmed arbitrary ultra-high-order mode operators can still maintain impressively well with the insertion loss less than 1.5 dB as well as crosstalk lower than -8.5 dB over an ultra-broad bandwidth of 150 nm, leading to a considerable progress to the de facto multimode photonics solutions.

## Results

**Mode operator building block.** Previously we have reported an ultra-compact mode-order converter by exploiting a fully etched silica slot on a silicon waveguide, which can efficiently serve as both a power splitter and a phase shifter at the same time [56]. Now we further develop this concept for universal mode operator including both manipulations of the conversation and exchange, utilizing the metasurface structures to realize the straightforward beam forming technique. We impose fully etched metasurface dielectric perturbations on silicon waveguides to split the input fundamental mode ($TE_0$) into $n+1$ beams with nearly equal power and simultaneously induce a $\pi$ phase difference between contiguous components. As a result, the $n_{th}$-order mode ($TE_n$) can be successfully formed at the end of the perturbations within several micrometers. In the following section, we take the $TE_0$-$TE_2$ mode operator as an example to illustrate the design methodology in detail and provide the theoretical analysis with the CMT model.

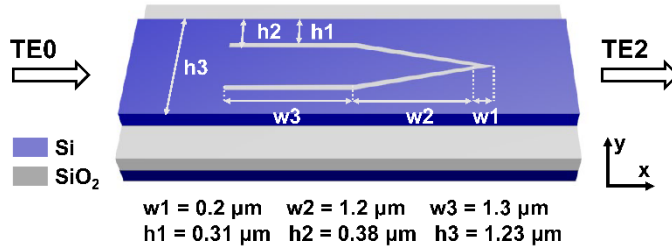

**Fig. 1** Schematic of the $TE_0$-$TE_2$ mode converter.

In the basic $TE_0$-$TE_2$ mode operator building block, the perturbation in the metasurface structure is initialized with a symmetric polygon shape as shown in Fig. 1. The two straight arms separate the multimode waveguide into three single-mode waveguides, where three $TE_0$-like components will transmit through. We sweep the variables in 3D finite-difference-time-domain (FDTD) simulations and the optimized parameters are given in Fig. 1. Fig. 2(a) presents the simulated electric field distribution at the wavelength of 1550 nm, which clearly shows the $TE_0$-$TE_2$ mode conversion. The simulated transmission spectrum is given in Fig. 2(b). The insertion loss is less than 1.5 dB from 1450 nm to 1600 nm, while the crosstalk is lower than -10 dB from 1400 nm to 1580 nm. The curve 'other' stands for the total crosstalk from other modes except for the $TE_0$ mode and the $TE_2$ mode.

Although the structure is originally intended for mode conversion, it can actually implement the function of mode exchange. To demonstrate this, we input TE$_2$ mode from the left of the structure and the simulation results are shown in Fig. 3. The TE$_0$ mode is eventually formed at the end of the device. Besides, the insertion loss is less than 1 dB and the crosstalk is lower than -10.5 dB from 1400 nm to 1600 nm, which is comparable to the performance when it serves as a mode converter.

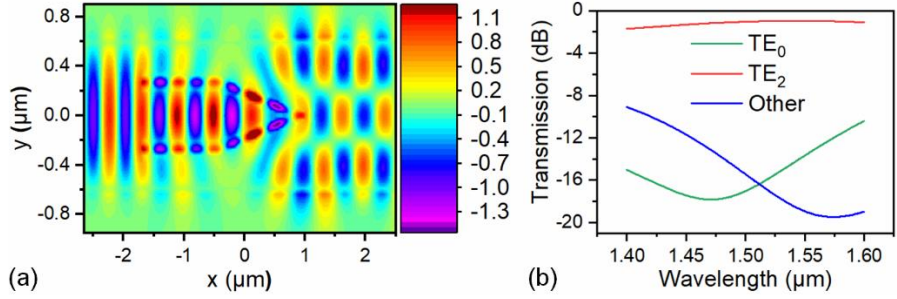

**Fig. 2** Simulated electric field distribution **(a)** and transmission spectra **(b)** of the TE$_0$-TE$_2$ mode converter

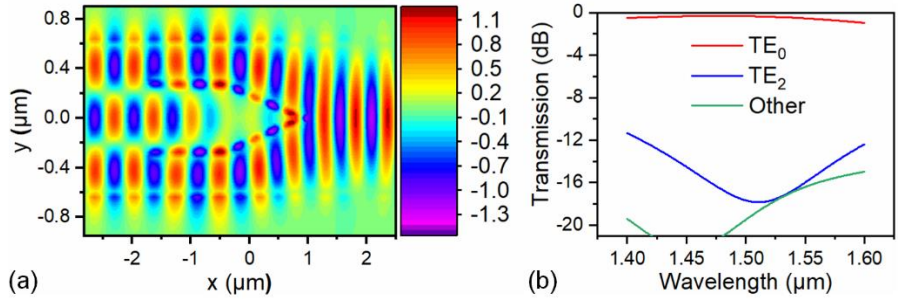

**Fig. 3** Simulated electric field distribution **(a)** and transmission spectra **(b)** of the TE0-TE2 mode exchanger.

**Arbitrary even high-order mode operators.** We now demonstrate that all even-order mode operators can be obtained by simply programming the same parallel perturbations with the TE$_0$-TE$_2$ mode operator building blocks. Although an individual building block can generate three channels, the two light beams on both sides actually have the same phase. To satisfy the anti-phase relation between neighboring beams, the distance between two adjacent building blocks has to be narrowed to generate only one channel. Consequently, total $2n+1$ channels can be generated by $n$ building blocks. Therefore, $n/2$ building blocks are needed to design a $n_{th}$-order ($n$ is an even number) mode operator, and the waveguide width can be expressed by

$$w_{even}=d\frac{n}{2}+2w_{extra}(\mu m) \tag{1}$$

where d is the central distance between adjacent building blocks, and $w_{extra}$ is the applied extra waveguide width to better confine the guided modes. The optimized parameters from FDTD simulations are $d=0.92$ μm and $w_{extra}=0.19$ μm, respectively.

As an example, the structure of the TE$_0$-TE$_6$ mode operator is presented in Fig. 4(a). Three same TE$_0$-TE$_2$ mode operators are programmed in a parallel array, and the whole structure is vertical symmetric along the propagation direction. Only seven TE$_0$-like channels are formed in the conversion region due to the narrow gaps between two adjacent building blocks. We first input the TE$_0$ mode to verify the function of TE$_0$-TE$_6$ mode converter. As depicted in Fig. 5 (a), total seven beams are progressively generated and eventually combine into the expected TE$_6$ mode. Figure 5(b) presents the simulated conversion efficiency and modal crosstalk. The insertion loss is measured to be less than 1.5 dB and the crosstalk is below -10 dB from 1430 nm to 1650 nm. Besides, the major crosstalk comes from the $TE_4$ mode. Then we input the $TE_6$ mode to further verify the function of $TE_0$-$TE_6$ mode exchanger. The electric field distribution depicted in Fig. 6(a) confirms the successful $TE_6$-$TE_0$ mode conversion. Meanwhile, the device exhibit encouraging performance with IL less than 1 dB as well as crosstalk lower than -10 dB from 1450 nm to 1650 nm as shown in Fig. 6(b). Therefore, the $TE_0$-$TE_6$ mode operator can function as a mode converter and a mode exchanger at the same time.

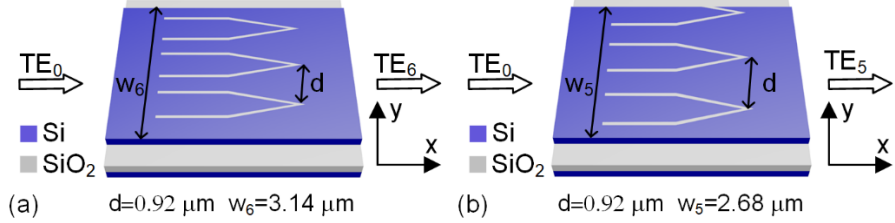

**Fig. 4** Schematic of the $TE_0$-$TE_6$ **(a)** and $TE_0$-$TE_5$ **(b)** mode operator.

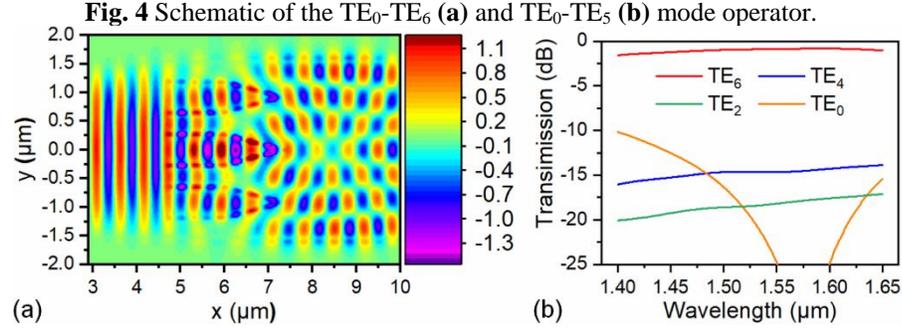

**Fig. 5** Simulated electric field distribution **(a)** and conversion efficiency **(b)** of the $TE_0$-$TE_6$ mode converter.

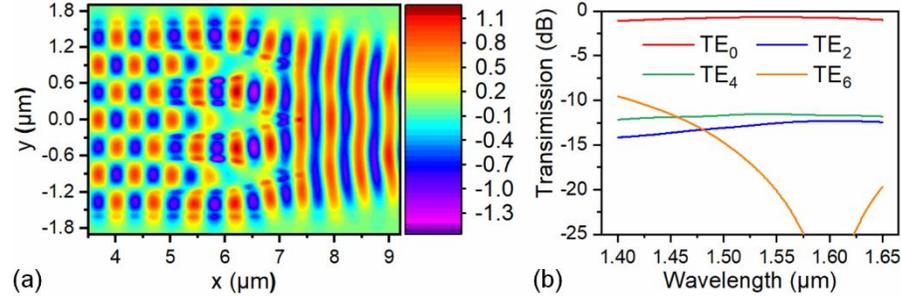

**Fig. 6** Simulated electric field distribution **(a)** and conversion efficiency **(b)** of the $TE_6$-$TE_0$ mode exchanger.

Actually, this approach can be extended for any even-order mode operators by programming the proper numbers of parallel building block and waveguide width as discussed above. As a proof of concept, we simulate an unprecedented high-order $TE_0$-$TE_{28}$ mode operator. As presented in Fig. 7(a), when we input the $TE_0$ mode, the insertion loss is still less than 1.5 dB and the modal crosstalk maintains lower than -8.5 dB from 1450 nm to 1600 nm. It's worth noting that the crosstalk mainly results from the neighboring guided mode $TE_{26}$, due to the similarity between their mode profiles. Although this is undesired and unavoidable, it may provide us with another simpler way to test the coupling efficiency of the ultra-high-order mode converters. Instead of quantifying the crosstalk form each single mode in traditional methods, we only focus on the specific modes causing the major crosstalk. Furthermore, the simulation results when we input the $TE_{28}$ mode are presented in Fig. 7(b). We emphasize that the performance maintains impressive well with IL less than 1.5 dB and crosstalk lower than 8.5 dB from 1420 nm to 1600 nm.

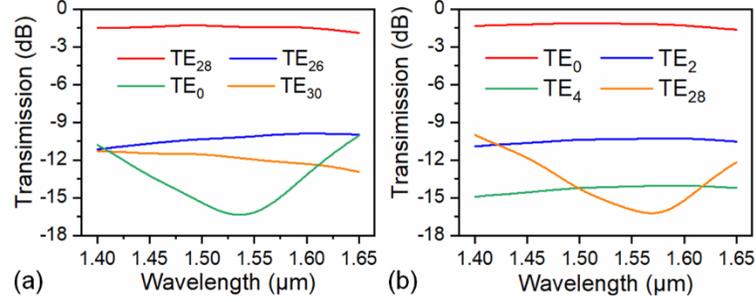

**Fig. 7** Simulated conversion efficiency of **(a)** the $TE_0$-$TE_{28}$ mode converter and **(b)** the $TE_{28}$-$TE_0$ mode exchanger.

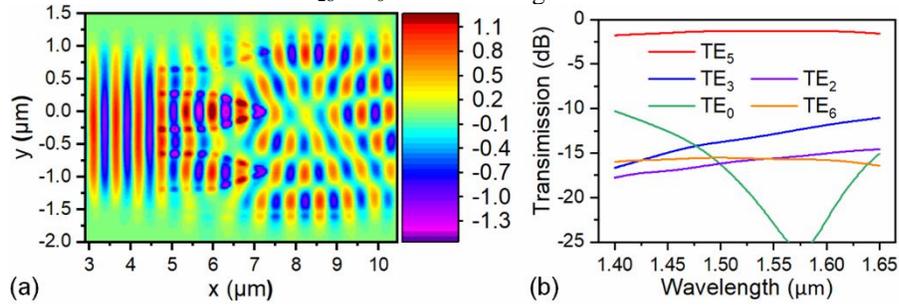

**Fig. 8** Simulated electric field distribution **(a)** and conversion efficiency **(b)** of the $TE_0$-$TE_5$ mode converter.

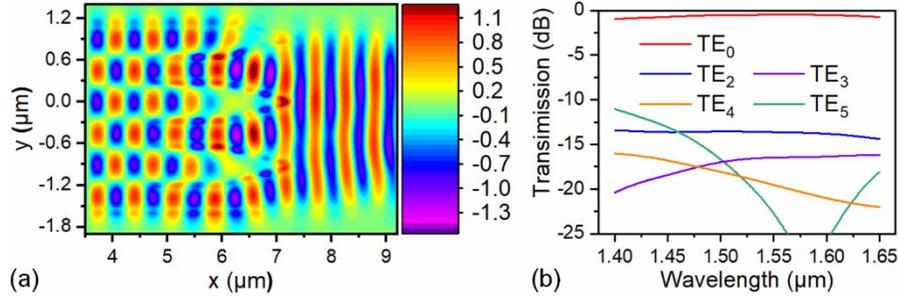

**Fig. 9** Simulated electric field distribution **(a)** and conversion efficiency **(b)** of the $TE_5$-$TE_0$ mode exchanger.

**Arbitrary odd high-order mode operators.** To further extend the generality of the proposed generic building block concept, we demonstrate the implementation of arbitrary high odd-order mode operators on the basis of even-order mode operators. To design a $n_{th}$-order ($n$ is an odd number) mode operator, we reduce one waveguide channel of the $TE_0$-$TE_{n+1}$ mode operator by directly chopping part of the structure on one side with a fixed offset, which is much smaller than the original waveguide width. Fortunately, the evolution of optical field will just be slightly influenced, thus remaining the desired $n$ single-mode channels. (This method is not suitable for the $TE_1$ and $TE_3$ mode, and their implementations based on the same design concept are given in Section 2, supporting information). Furthermore, the waveguide width for odd-order mode operators can be expressed by

$$w_{odd}=d\frac{n+1}{2}+2w_{extra}-w_{offset}\ (\mu m) \quad (2)$$

where $w_{offset}$ is the width of the truncated part, and the optimized value is 0.46 μm.

As an example, the schematic diagram of the $TE_0$-$TE_5$ mode operator is shown in Fig. 4(b), which is exactly the same as that of the $TE_0$-$TE_6$ mode operator, except 0.46 μm wide structure is cut off on the upper side. The simulated electric field distribution and the transmission curves for both the mode converter and the mode exchanger are given in Fig. 8 and Fig. 9 respectively. The insertion loss is less than 1.5 dB from 1450 nm to 1620 nm in both situations. Besides, the crosstalk is below -10 dB when we input the $TE_0$ mode and even lower than -12.5 dB when we input the $TE_5$ mode. Likewise, we give the simulation results for the ultra-high $TE_0$-$TE_{27}$ mode operators to show the scalability in Fig. 10. Similarly, the insertion loss keeps less than 1.5 dB and the crosstalk still maintains lower than -8.5 dB from 1430 nm to 1600 nm. Besides, as shown in the Fig. 10(a), the $TE_{25}$ mode accounts for the major crosstalk for the mode converter.

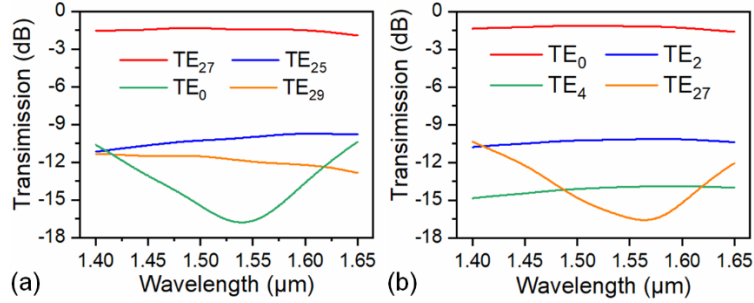

**Fig. 10** Simulated conversion efficiency of **(a)** the $TE_0$-$TE_{27}$ mode conversion and **(b)** the $TE_{27}$-$TE_0$ mode conversion.

## Discussion

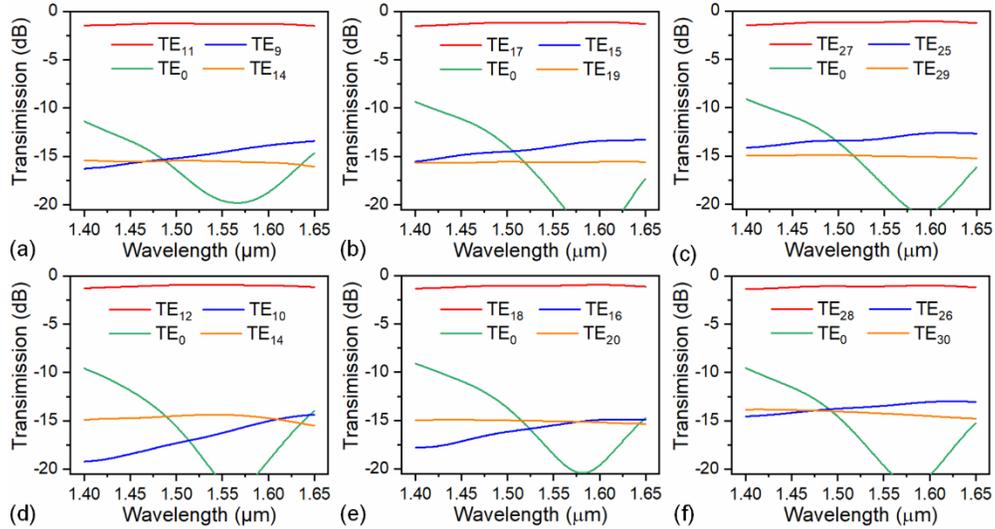

**Fig. 11** Simulated conversion efficiency of the $TE_0$-$TE_{11}$ **(a)** $TE_0$-$TE_{17}$ **(b)** $TE_0$-$TE_{27}$ **(c)** $TE_0$-$TE_{12}$ **(d)** $TE_0$-$TE_{18}$ **(e)** $TE_0$-$TE_{28}$ **(f)** mode converter.

Due to the intrinsic Gaussian field distribution of the $TE_0$ mode, the energy on both sides of the waveguide will decrease much smaller with the increase of the mode order, which directly pose limitations on the beam forming performance of the proposed parallel layouts. Consequently, the modal crosstalk for high-order mode converters is limited to be approximately -8.5 dB. To alleviate this problem, a taper guiding the input $TE_0$ mode from a broader waveguide can be added before the programmable structures presented above. The optimized taper length is 6 μm and the angle is 12 degrees for all the high-order mode converters (The performance analysis of the designed tapers is provided in Section 3, supporting information). The simulation results for the $TE_0$-$TE_{17}$, $TE_0$-$TE_{18}$, $TE_0$-$TE_{27}$ and $TE_0$-$TE_{28}$ mode converters are presented in Fig. 11. Obviously, the performance is remarkably improved with the crosstalk lower than -12 dB from 1450 nm to 1650 nm for all the six occasions. We noted that the taper won't work for the mode exchangers, considering that high-order modes will evolve into various different modes after transmitting through the short tapers.

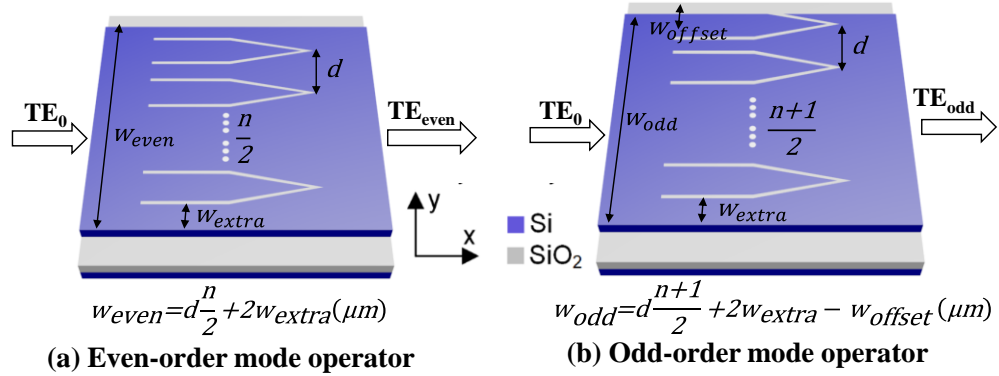

$$w_{even}=d\frac{n}{2}+2w_{extra}(\mu m)$$

**(a) Even-order mode operator**

$$w_{odd}=d\frac{n+1}{2}+2w_{extra}-w_{offset}(\mu m)$$

**(b) Odd-order mode operator**

**Fig. 12** Universal arbitrary high-order mode operator: **(a)** even modes **(b)** odd modes

A novel universal design for arbitrary high-order mode operators based on the programmed metasurface building blocks on silicon has been demonstrated and illustrated in Fig. 12. The estimated over-all performance of the programmed building-block-implemented arbitrary mode operators is as follows: The insertion loss is predicted to be approximately 1.5 dB. Besides, the neighboring $TE_{n-2}$ mode is the major source of the modal crosstalk for the $TE_0$-$TE_n$ mode conversion, which is estimated to be nearly -8.5 dB within a broad bandwidth from 1450 nm to 1600 nm. Moreover, the crosstalk of high-order mode converters can be further enhanced to be lower than -12 dB by utilizing properly designed tapers. We emphasize that the footprint of the proposed mode operator increases linearly with the mode order, promising high integration intensity. Since we just engineer the building blocks in a generic rough way, the acquired decent performance can be further improved by coarsely tuning the parameters (e.g., the waveguide width, the distance between building blocks) for specific mode operators. Moreover, the mode coupling efficiency can be flexibly tailored by adjusting the arm lengths of the building blocks.

The proposed programmable building block concept can be efficiently migrated to other platforms (e.g., InP, $Si_3N_4$, etc.) as well as other wavelength-bands (e.g., mid-infrared band, etc.), which will definitely open the opportunity to better control the multimode photonics on-chip. The simultaneous manipulation of multiple modes will not only increase the link capacity in optical communication systems, but also potentially find applications in quantum information processing [57], nonlinear photonics [58], photonic sensing [59], etc.

In conclusion, by employing the programmable compact metasurface building blocks, we report a universal methodology to realize the mode conversion and mode exchange between the fundamental mode and arbitrary high-order mode for the first time. All the even-order and odd-order mode operator can be efficiently realized by programming the parallel array of the building block, i.e., the $TE_0$-$TE_2$ mode operator and coarsely control the waveguide widths. The proposed devices feature constant performance of broad bandwidth (from 1450 nm to 1600 nm), low insertion loss (<1.5 dB), low modal crosstalk (-8.5 dB), compact footprints and robustness to the fabrication variations. Moreover, the crosstalk of the mode converters can be improved to be lower than -12 dB by adding a taper. This will give significant inspirations to other device designs and could promise a great breakthrough to boost the development of on-chip MDM photonic systems.

## Methods.
### Coupled mode theory.

The evolution of optical field in a perturbed structure can be described by the CMT model (see the detailed analysis in Section 1, supporting information). Fig. 13(a) presents the mode purity from both numerical calculations of CMT model (curves) and FDTD simulation (symbols) along the propagation direction, which agree quite well with each other. It's clear that the input $TE_0$ mode is gradually converted into the $TE_2$ mode within a short length less than 3 μm. Fig. 13(b) shows the coupling coefficients as a function of the propagation distance, which is not sinusoidal-like owing to the aperiodic perturbations. It's necessary for $\kappa_{02}$ to change from negative to positive value, ensuring that the $TE_0$ mode always contributes constructively to the conversion of the $TE_2$ mode. Fortunately, there is almost zero crosstalk from the $TE_1$ mode due to the negligible $\kappa_{01}$.

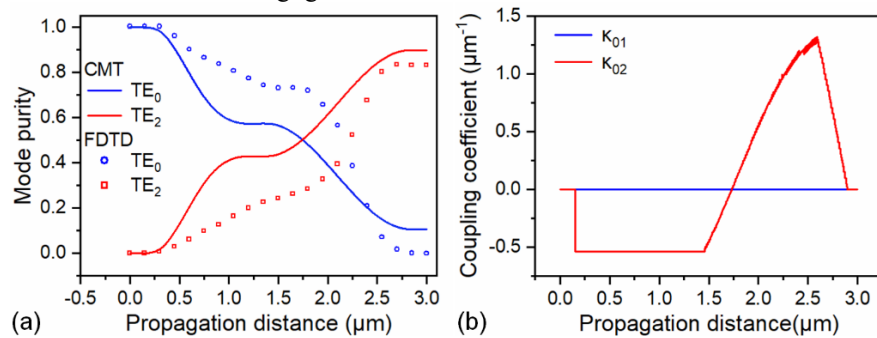

**Fig. 13** Mode evolution **(a)** and coupling coefficients **(b)** as a function of propagation distance.

### Device fabrication.

### Measurement setup.

**Funding.** National Key R&D Program of China (2019YFB2203101); Natural Science Foundation of China (NSFC) (61805137 and 61835008); Natural Science Foundation of Shanghai (19ZR1475400); Shanghai Sailing Program (18YF1411900); Open Project Program of Wuhan National Laboratory for Optoelectronics (2018WNLOKF012).

**Disclosures.** The authors declare no conflicts of interest.